\documentclass[iop,revtex4]{emulateapj} 
\usepackage{amsmath}  
\usepackage[titletoc,title]{appendix}
\usepackage{graphicx,natbib,url,twoopt}
\usepackage[varg]{txfonts}
\usepackage{hyperref}               %% for pdflatex
\usepackage{blkarray}

\bibpunct{(}{)}{;}{a}{}{,}    %% natbib cite format used by A&A and ApJ

\received{receipt date}
\revised{revision date}
\accepted{acceptance date}
\begin{document}  

%% simple header.  Change into A&A or ApJ commands for those journals
\title{Measuring the microlensing parallax from various space observatories}

\author{E. Bachelet$^{1}$, T. C. Hinse$^{2}$ and R. Street$^{1}$}

\affil{$^{1}$Las Cumbres Observatory, 6740 Cortona Drive, Suite 102, Goleta, CA 93117 USA}
\affil{$^{2}$Korea Astronomy \& Space Science Institute, 305-348 Daejeon, Republic of Korea}
\begin{abstract}
A few observational methods allow the measurement of the mass and distance of the lens-star for a microlensing event. A first estimate can be obtained by measuring the microlensing parallax effect produced by either the motion of the Earth (annual parallax) or the contemporaneous observation of the lensing event from two (or more) observatories (space or terrestrial parallax) sufficiently separated from each other.
Further developing ideas originally outlined by \citet{Gould2013b} and \citet{Mogavero2016}, we review the possibility of measuring systematically the microlensing parallax using a telescope based on the Moon surface and other space-based observing platforms including the upcoming WFIRST space-telescope. We first generalize the Fisher matrix formulation and present results demonstrating the advantage for each observing scenario. We conclude by outlining the limitation of the Fisher matrix analysis when submitted to a practical data modeling process. By considering a lunar-based parallax observation we find that parameter correlations introduce a significant loss in detection efficiency of the probed lunar parallax effect.
\end{abstract}
\keywords{gravitational microlensing-parallaxes}
%%%%%%%%%%%%%%%%%%%%%%%%%%%%%%%%%%%%%%%%%%%%%%%%%%%%%%%%%%%%%%%%%%%%%%%%%%%%
\section{Introduction}     \label{sec:introduction}
%%%%%%%%%%%%%%%%%%%%%%%%%%%%%%%%%%%%%%%%%%%%%%%%%%%%%%%%%%%%%%%%%%%%%%%%%%%%
Measuring the microlensing parallax is of primary importance, since it constrains 
the mass-distance relation of the microlensing lens and allows the physical properties of the lens to be measured \citep{Gould2000}:
\begin{equation}
M_l =  {{\theta_E}\over{\kappa\pi_E}}
\end{equation}
where $M_l$ is the lens mass in solar unit, $\theta_E$ is the angular Einstein ring radius,  $\pi_E$ is the microlensing parallax and $\kappa=8.144~ \rm{mas.M_\odot^{-1}}$. The microlensing parallax can be measured in three possible observing scenarios. 
The non-rectilinear motion of the Earth around the Sun imposes an additional kinematic component on the 
relative lens-source trajectory $\mu$ and is known as the annual parallax effect \citep{Alcock1995,Gould2000,Smith2003, Gould2004}.
The effect is greater for long event timescales and typically events with an angular Einstein ring crossing time  $t_E=\theta_E/\mu \ge 30$ days present significants variations. 
This effect is also greater when the observations occurs near the equinoxes \citep{Skowron2011}. The second method requires the microlensing event 
to be observed from two observatories separated by a significant baseline, see for example \citet{Refsdal1966, CalchiNovati2015,Street2016,Henderson2016}. This method is called the space parallax, since it generally involves the use 
of ground and space-based observatories.
The last method, called the terrestrial parallax, is hard to measure, but has been measured in few cases \citep{Yee2009,Gould2009}. The separation between 
two distinct observatories on Earth, with different location in longitude and latitude, induces a shift in both the time of event magnification $t_0$ and minimum impact parameter $u_0$. Since the separation is small relative to the projected Einstein radius, this 
effect is measurable only for extreme high magnification events \citep{Hardy1995,Holz1996,Gould1997,Gould2013a}.

Recently, \citet{Gould2013b} and \citet{Mogavero2016} (thereafter G13 and M16) 
explored the capability to measure parallax using space-based observatories only.
They concluded that this is feasible for observing platforms on geosynchronous and Low Earth Orbits (LEO), 
depending on the specific microlensing event signal-to-noise ratio. Their work motivated us to study this aspect of parallax measurements in some more detail. 

The outline of this work is as follows. We extend the 
approach of G13 and M16 and worked out a more general description of the Fisher matrix formulation in Section~\ref{sec:Fisher}. In Section~\ref{sec:single} 
and Section~\ref{sec:fleet}, we then study the potential of a wider range of space-based observatories in order to measure the microlensing 
parallax. In Section~\ref{sec:real}, we highlight the difficulty to detect parallax in practice. This is a consequence of the parallax being an observable obtained from a best-fit model that suffers from correlations between parameters. We conclude our study in Section~\ref{sec:conclusions}.
%%%%%%%%%%%%%%%%%%%%%%%%%%%%%%%%%%%%%%%%%%%%%%%%%%%%%%%%%%%%%%%%%%%%%%%%%%%%
\section{Parallax formulation and Fisher matrix analysis} \label{sec:Fisher}
%%%%%%%%%%%%%%%%%%%%%%%%%%%%%%%%%%%%%%%%%%%%%%%%%%%%%%%%%%%%%%%%%%%%%%%%%%%%
%%%%%%%%%%%%%%%%%%%%%%%%%%%%%%%%%%%%%%%%%%%%%%%%%%%%%%%%%%%%%%%%%%%%%%%%%%%%
\subsection{Parameterization of the problem} 
\begin{figure}    
  \centering
  \includegraphics[width=8cm]{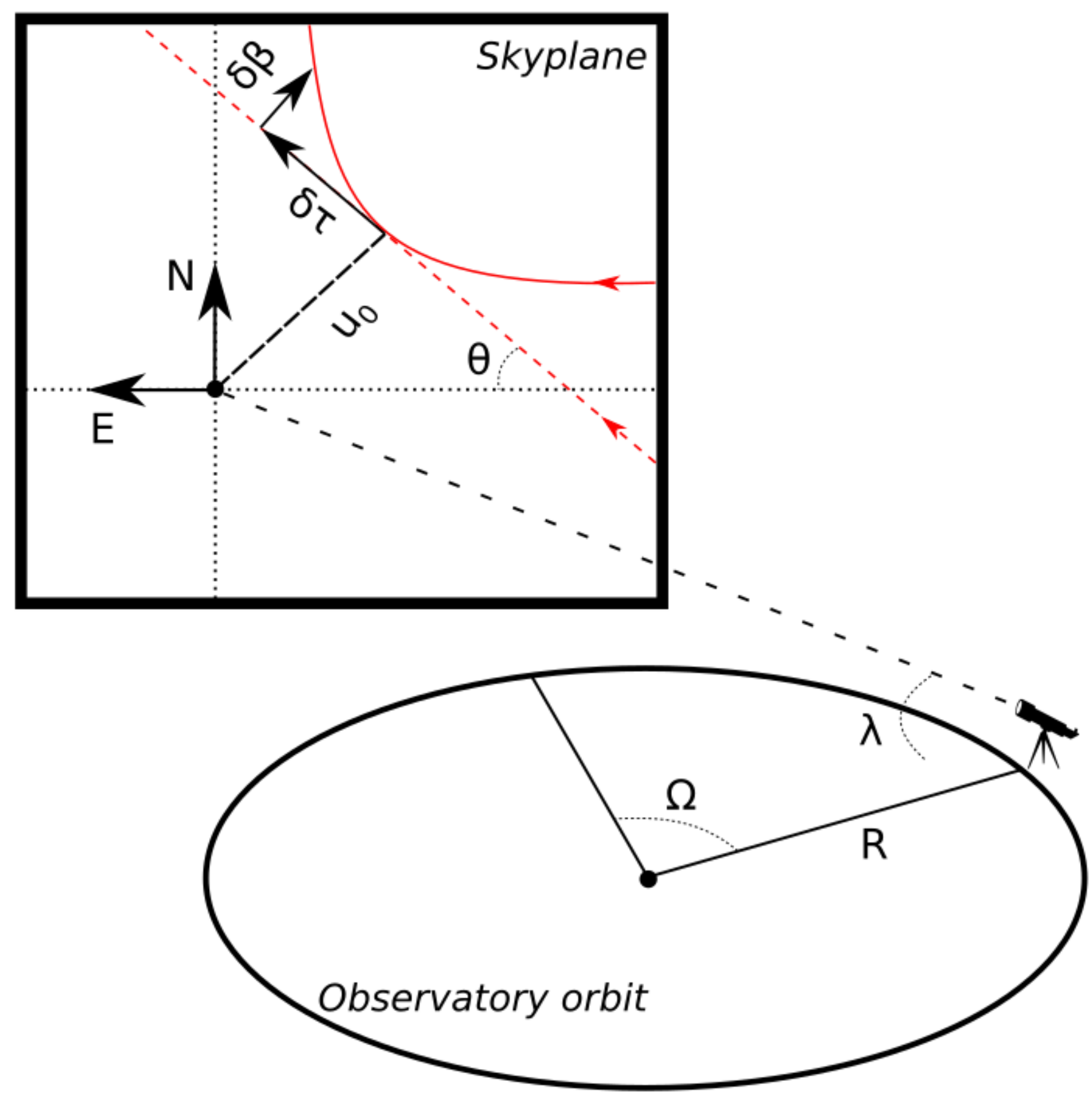}
    \caption{Schematic representation of the problem. As the observatory travels in its orbit, the source trajectory (solid red) is shifted from the inertial trajectory (dash red). The position of the lens is indicated by a point in the skyplane. $(\delta\tau,\delta\beta)$ are represented at the time $t_0$.}
    \label{fig:geometry}
\end{figure}

%%%%%%%%%%%%%%%%%%%%%%%%%%%%%%%%%%%%%%%%%%%%%%%%%%%%%%%%%%%%%%%%%%%%%%%%%%%%
Following the method outlined in G13 and M16, we conduct a Fisher matrix analysis
for various space-based observatories. G13 and M16 consider observatories
with orbital radii that are small compared with 1 AU, which allows some approximations
in the Fisher matrix analysis. This approximation can not be applied to the present work because we consider observatories separated by 
a large orbital radius, for example a satellite orbiting the Sun at 1 AU. Therefore, a general 
Fisher matrix analysis is required.
For simplicity, we consider only circular orbits in this work without loss of generality. As in G13 and M16, we first define $\epsilon_\parallel=\rm{R}/\rm{AU}$ and $\epsilon_\bot=\epsilon_\parallel\sin{\lambda}$, where
$R$ is the the orbital radius of the observatory platform (associated with a period $P$) and $\lambda$ is the latitude of
the microlensing target relative to the observatory orbital plane. If we consider the problem in the reference frame centered on the observatory at the microlensing peak $t_0$, the motion of the coordinates of the observatory $\mathbf{O} = (o_1,o_2)$ are:
\begin{equation}
\begin{aligned}
o_1 &=& \epsilon_\parallel\cos{\Omega}-\epsilon_\parallel\cos{\phi} \\ 
o_2 &=& \epsilon_\bot\sin{\Omega}-\epsilon_\bot\sin{\phi}
\end{aligned}
\label{eq:positions}
\end{equation}
with $\Omega = \omega(t-t_0)+\phi$ , $\omega = 2\pi/P$ and $\phi$ is the orbital phase relative to microlensing event time of maximum magnification. This approach is similar to that of \citet{Gould2004}. We now define $\tau = (t-t_0)/t_E$, $u_0$ and $\theta$ (the lens-source trajectory angle) as the standard microlensing parameters for the static observatory (see for example \citet{Gould2000} for the definition of these parameters, as well as the Figure~\ref{fig:geometry}). If one defines the microlensing parallax vector as $\mathbf{\pi_E} = (\pi_\parallel,\pi_\bot) = \pi_E(\cos{\theta},\sin{\theta})$, the moving observatory $(\delta\tau,\delta\beta)$ shifts are:
\begin{equation}
\begin{aligned}
\delta\tau &=&  \mathbf{\pi_E} \cdot  \mathbf{O}\\
\delta\beta &=& \mathbf{\pi_E} \times \mathbf{O}\\
\end{aligned}
\end{equation}
Defining $\tau' = \tau+\delta\tau$ and $\beta = u_0+\delta\beta$, the microlensing trajectory vector is then $\mathbf{u} = (u_1 = \tau'\cos{\theta}-\beta\sin{\theta} ~,~  u_2 = \tau'\sin{\theta}+\beta\cos{\theta})$. The observed flux of the lensing event is:
\begin{equation}
f = f_s(A+g)
\end{equation}
with $f_s$ the source flux and $g=f_b/f_s$ is the blending ratio ($f_b$ is the blend flux). The source flux magnification $A(t)$ for a single point lens is a function of time and is given by \citep{Paczynski1986} :
\begin{equation}
A(t) = {{u(t)^2+2}\over{u(t)\sqrt{u(t)^2+4}}}
\end{equation}
where $u(t) = \sqrt{u_1^2+u_2^2}$.

We follow M16 and \citep{Bachelet2017} and, assuming Gaussian errors,  we define the Fisher matrix as :
\begin{equation}
F_{i,j} =\sum_{n}{{1}\over{\sigma_n^2}}{{dF_n}\over{dp_i}}{{dF_n}\over{dp_j}}
\end{equation}
where $n$ indicates the number of measurements. Here, we follow M16's approach and eliminate the source flux from the flux derivatives ${dF_n}\over{dp_i}$ of the Fisher matrix and from the weight with $\sigma_n^2 \simeq 0.84 \sigma_m^2 (A+g)/(1+g)$, where $\sigma_m$ is an arbitrary 
photometric precision (in magnitude units) for the microlensing event baseline magnitude. The individual 
derivatives can be found in Appendix~\ref{sec:derivatives}. The covariance matrix is then simply the inverse of the Fisher matrix:
\begin{equation}
\rm{cov} =F^{-1}
\end{equation}
M16 defines the minimum error on the parallax measurement $\sigma_{\pi_E, min}(\phi)$ as:
\begin{equation}
\sigma ^2_{\pi_{E}, min}(\phi)= {{\sigma_{\pi_\parallel}^2+\sigma_{\pi_\perp}^2}\over{2}} - {{\sqrt{(\sigma_{\pi_\parallel}^2-\sigma_{\pi_\perp}^2)^2+4~\rm{cov}(\pi_\parallel,\pi_\perp)^2}}\over{2}}
\end{equation}
As a sanity check, we compare our estimation of $\sigma_{\pi_E,min}(\phi)$ with the one 
found by M16 for the case of a geosynchronous observatory, assuming P = 23h 56min 4s, $\rm{R} =6.6~\rm{R}_\oplus$, $\lambda=30^{\circ}$, $u_0=0.1$, $t_E=1$day, $\theta=45^{\circ}$, $\pi_E=4.3$, $g=0$, $\sigma_m=0.01$ mag, $\phi=0$, 180 days of observation around $t_0$ and an observing cadence of 3 min. M16 Fisher matrix formulation leads to $\sigma_{\pi_E,min}(0)/\pi_E\sim 0.08$ and our estimation gives a good agreement of $\sigma_{\pi_E,min}(0)/\pi_E\sim 0.06$.

%%%%%%%%%%%%%%%%%%%%%%%%%%%%%%%%%%%%%%%%%%%%%%%%%%%%%%%%%%%%%%%%%%%%%%%%%%%%
\subsection{Hypothesis and assumptions} 
%%%%%%%%%%%%%%%%%%%%%%%%%%%%%%%%%%%%%%%%%%%%%%%%%%%%%%%%%%%%%%%%%%%%%%%%%%%%
For the remainder of this paper, we will study the microlensing parallax measurement for observatories
orbiting the Sun (Section~\ref{sec:fleet}), the Earth (Section~\ref{sec:moon}) and the Lagrangian point L2 (Section~\ref{sec:wfirstpotential}). In principle, the change in the origin of the reference system to each of these locations should be taken into account. However, this introduces considerable additional complexity into the Fisher matrix derivation, so for the time being we neglect the impact of the inertial reference point, which is a valid approximation for event timescales which are short compared with the orbital period of the inertial reference point. Note that both G13 and M16 also neglect this effect. 
We consider events whose photometry is not blended with the light from neighboring (unrelated) stars (i.e $g=0$). Note that the authors in M16 stressed that blending can have a serious effect on the parallax detection. We also consider continuous observations to reduce complexity. M16 indicates that while the Earth's umbra effectively decreases the sensitivity of LEO satellites, it does not invalidate the method. We assume Keplerian orbits, so the period of our observatories is obtained from Kepler's law $P^2 = 4\pi^2/(GM)R^3$, the mass depending on the system considered. We also assume Gaussian errors due to the nature of space-based observations. Finally, throughout this study, we assume that the source is located in the Galactic Bulge (i.e $D_s=8$ kpc), the lens is located at $4$ kpc and a relative source-lens speed $V=200$ km/s, leading to :
\begin{equation}
\pi_E = 4.3 ~\bigg({{\rm{1~day}}\over{t_E}}\bigg)
\end{equation}

%%%%%%%%%%%%%%%%%%%%%%%%%%%%%%%%%%%%%%%%%%%%%%%%%%%%%%%%%%%%%%%%%%%%%%%%%%%%
\section{Single observatory} \label{sec:single}
%%%%%%%%%%%%%%%%%%%%%%%%%%%%%%%%%%%%%%%%%%%%%%%%%%%%%%%%%%%%%%%%%%%%%%%%%%%%
%%%%%%%%%%%%%%%%%%%%%%%%%%%%%%%%%%%%%%%%%%%%%%%%%%%%%%%%%%%%%%%%%%%%%%%%%%%%
\subsection{The parallax seen from the Moon} \label{sec:moon}
%%%%%%%%%%%%%%%%%%%%%%%%%%%%%%%%%%%%%%%%%%%%%%%%%%%%%%%%%%%%%%%%%%%%%%%%%%%%
In the following we consider the case of a single telescope based on the surface of the Moon. For a higher sky visibility, the preferable observatory location is on the lunar dark side, but could raise practical difficulties, especially communications. The Earth-facing lunar hemisphere seems to be more practical. This is the choice made by the China National Space Administration to place the first robotic telescope 
on the Moon \citep{Wang2015b}. The 15 cm diameter Lunar-based Ultraviolet Telescope currently operates from the \textit{Mare Imbrium} 
with a photometric precision of $\sigma\sim 0.05$ for a $\sim 17.5$ mag star (in AB photometric system) \citep{Wang2015b}.
To understand the power of a lunar-based observatory with respect to microlensing parallax measurements, we consider the orbit of the Moon around the Earth to be circular, with an orbital radius $R=381600$ km and a photometric precision of $\sigma_m =  0.01$ mag for the event baseline magnitude.
We select $\log_{10}(u_0) \in [-5,0.3] $ and $\log_{10}(t_E) \in [-1,2]$. This parameter range is typical for microlensing events observed in the Galactic Bulge. We also consider $\theta=45^{\circ}$, $\phi=0^{\circ}$ and $\lambda = 35 ^{\circ}$. The last assumption comes from the fact that the
Moon's orbital inclination to the ecliptic plane is roughly $5^{\circ}$. 
Finally we construct the observing strategy as follows. We assume the lunar telescope observes a given event during two observing windows 
separated by a time interval of P (i.e $\sim$ 28 days). Each window consists of 14 days of continuous observations with 15 min sampling.
The first observing window is centered on $t_0$. The aim now is to calculate the minimum parallax error as a function of $u_0$ and $t_E$ from a general Fisher matrix formulation. 

\begin{figure}    
  \centering
  \includegraphics[width=9cm]{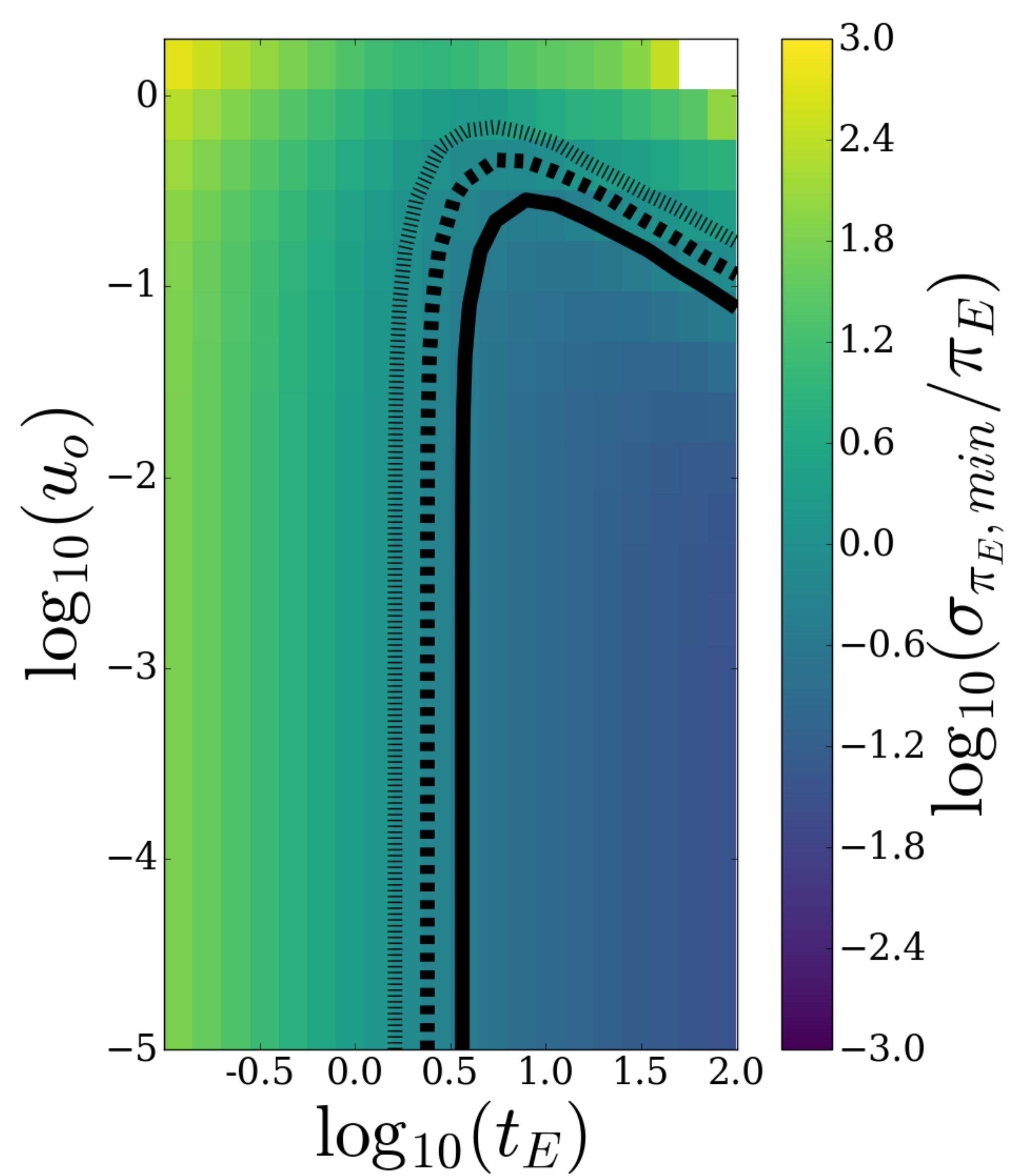}
    \caption{Minimum expected error on the parallax measurement $\sigma_{\pi_E,min}/\pi_E$ (color coded in $\log_{10}$ scale, in the range [-3,3]), for a telescope placed on the Moon. The small-dashed, dashed and solid contour curves indicate the 1, 2 and 3 $\sigma$ detection regions. The blank pixels on the top right indicates ill-observed event, leading to $\sigma^2_{\pi_E,min}(0)<0$.}
    \label{fig:Fisher1}
\end{figure}

Results of our simulations can be seen in Figure~\ref{fig:Fisher1}. Similarly to M16, the relative error is separated in two regimes, $u_0t_E<<P$ and $u_0t_E>>P$. From the figure we find that long timescale events ($t_E>40$ days) are ideal to securely estimate the associated parallax effect well within the $3 \sigma$ detection limit.  In general, lunar-based parallax measurements with errors less than $3 \sigma$ have $t_E > 10$ days. For $\sigma_{\pi_E,min}/\pi_E>3$, the parallax estimate is less well constrained, corresponding to events with timescales shorter than $t_E<5$ days. Such events could be caused by free-floating planets \citep{Sumi2010,Mroz2017}. Given that the (Galactic Bulge) microlensing timescale distribution peaks around $t_E\sim20$ days \citep{Sumi2010,Mroz2017}, we conclude that a dedicated microlensing monitoring telescope placed on the Moon could provide a valuable observing platform for the systematic and accurate sampling of most microlensing parallax measurements.
%%%%%%%%%%%%%%%%%%%%%%%%%%%%%%%%%%%%%%%%%%%%%%%%%%%%%%%%%%%%%%%%%%%%%%%%%%%%
\subsection{WFIRST} \label{sec:wfirstpotential}
%%%%%%%%%%%%%%%%%%%%%%%%%%%%%%%%%%%%%%%%%%%%%%%%%%%%%%%%%%%%%%%%%%%%%%%%%%%%
In this section we carry out a similar study considering NASA's WFIRST space satellite mission, which
will survey the Galactic Bulge in the near-infrared, with six observing windows 
of $\sim$ 70 days \citep{Spergel2015}. Contrary to the assumptions of G13 and M16, it has recently
been decided that WFIRST will be placed in a so-called halo-orbit at the Lagrangian point L2. This location offers many operational 
benefits, see for example \citet{Crowley2016}. 
 It is likely that the 
orbital elements of WFIRST will be similar to the Lissajous orbits of GAIA \citep{Perryman2001} and Planck \citep{Tauber2010,Pilbratt2010}. An L2 halo orbit has a relatively long period $P\sim180$ days and a orbital radius of few percent of an AU.
Following \citet{Henderson2016}, we consider WFIRST orbital parameters similar to 
the GAIA space mission : $P=180$ days and $R=300000$ km. For clarity, we assume $P$ to be the orbital period of WFIRST 
around the unstable Lagrangian point L2 (i.e. we did not include the movement of L2 around the Sun due to the motion of the Earth) at a fix distance $R$. 
We choose $\theta=45^{\circ}$, $\phi=0^{\circ}$, $\lambda = 30 ^{\circ}$
and set the monitoring window to 70 days centered around the peak magnification with a 15 min observing cadences. We follow M16 and G13 and assume a photometric precision $\sigma_m =  0.01$ mag as well as $\sigma_m =  0.001$ mag.
\begin{figure}    
  \centering
  \includegraphics[width=9cm]{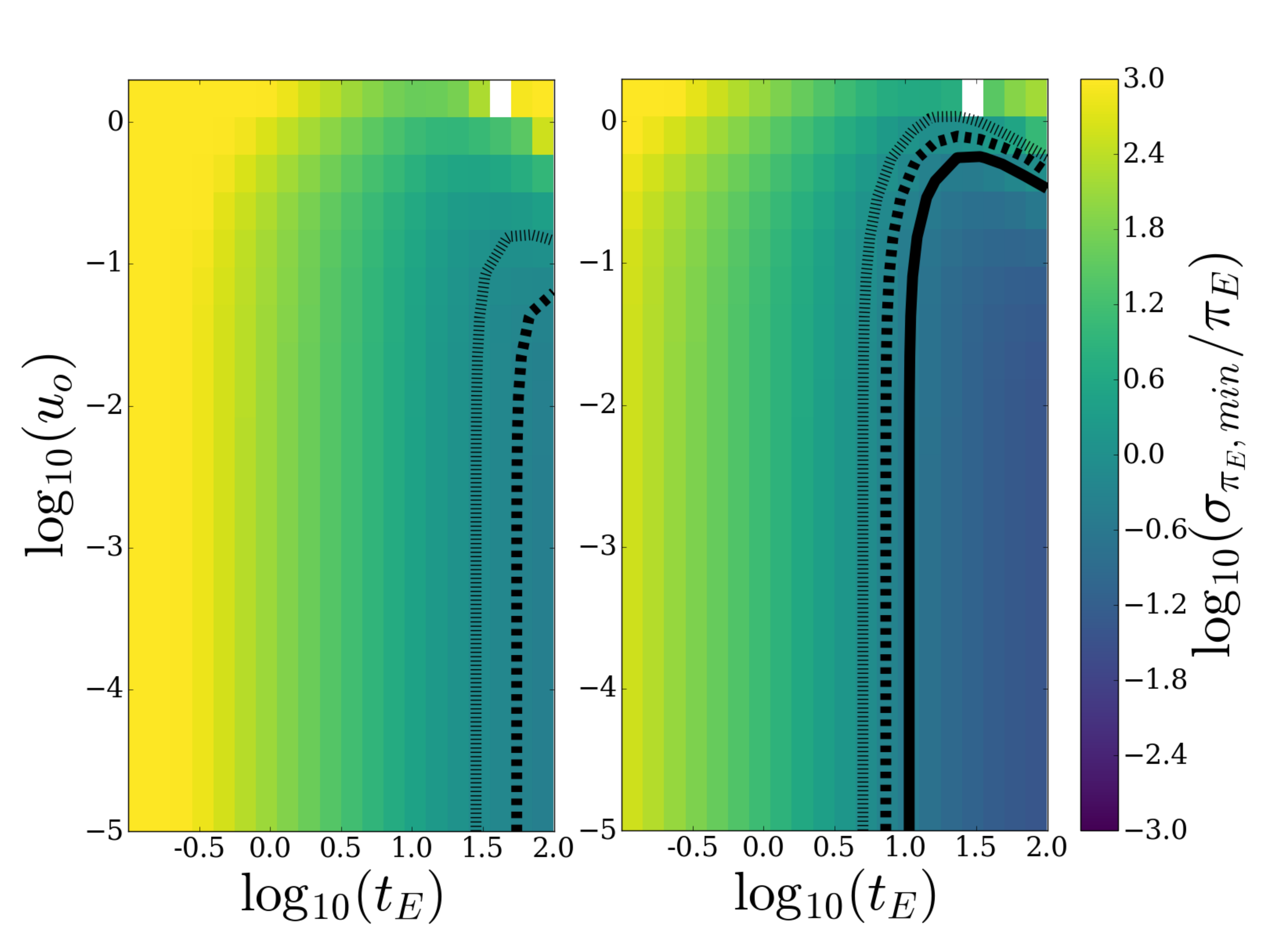}
    \caption{Same as Figure~\ref{fig:Fisher1}, for the WFIRST mission with observation parameters detailed in the text. \textit{Left:} Using $0.01$ mag photometric precision. \textit{Right:} Using $0.001$ mag photometric precision.}
    \label{fig:Fisher2}
\end{figure}

In Figure~\ref{fig:Fisher2} we show the minimum parallax error for two photometric precisions attributed to WFIRST platform specifications.
From the left panel, for a photometric precision of 0.01 mag, we find that WFIRST is not suitable to reliably measure the parallax which is explained by the long (L2 halo) orbital period. The minimum parallax error is given by (see M16):
\begin{equation}
\sigma_{\pi_E,min}/\pi_E \propto P^{0.5}u_0^{0.5}R^{-1}
\end{equation}
For brighter lensing events the photometric precision increases which could decrease the minimum parallax error. From the Fisher matrix formulation, we have therefore calculated the minimum parallax error for a photometric precision of 0.001 mag. The results are shown in the right panel of Figure~\ref{fig:Fisher2} and demonstrate that WFIRST is capable of measuring the event parallax for event timescales $t_E > 10$ days.
It is important to recall that we consider the Lagrangian point L2 stationary during WFIRST observations. This hypothesis breaks for longer events, where the combination of the two movements can in fact constrain the parallax well, see G13. Moreover, contemporaneous observations from WFIRST and ground-based observatories will allow the 
measurement of the so called space-based parallax \citep{Refsdal1966, CalchiNovati2015,Street2016,Henderson2016}.
However, these follow-up observations from ground could be challenging, due to potential high-extinction fields (that require 
near infrared observations) and/or low overlap between the observability windows from Earth and L2 observatory.
%%%%%%%%%%%%%%%%%%%%%%%%%%%%%%%%%%%%%%%%%%%%%%%%%%%%%%%%%%%%%%%%%%%%%%%%%%%%
\section{The parallax from a telescope constellation } \label{sec:fleet}
%%%%%%%%%%%%%%%%%%%%%%%%%%%%%%%%%%%%%%%%%%%%%%%%%%%%%%%%%%%%%%%%%%%%%%%%%%%%
Telescope constellation of small satellites, such as NASA CubeSat, is a relatively new and low-cost technology that could be competitive with fewer and larger satellites in the future. Here we consider a fleet of space telescopes in various orbital configurations. Since we consider several observatories, 
we need to choose a common origin. We define the origin of the system as the center of the trajectories. Then, the problem definitions are slightly changed and Equation~\ref{eq:positions} becomes:
\begin{equation}
\begin{aligned}
o_1 &=& \epsilon_\parallel\cos{\Omega}\\ 
o_2 &=& \epsilon_\bot\sin{\Omega}
\end{aligned}
\end{equation}
This implies that the microlensing parameters refer now to this origin ($u_0$ and $t_0$ especially), but the Fisher matrix formalism is unchanged 
since we subtracted constants. This is similar to the heliocentric and geocentric approaches for the annual parallax, see \citet{Gould2004}.

\begin{figure*}    
  \centering
  \includegraphics[width=\textwidth]{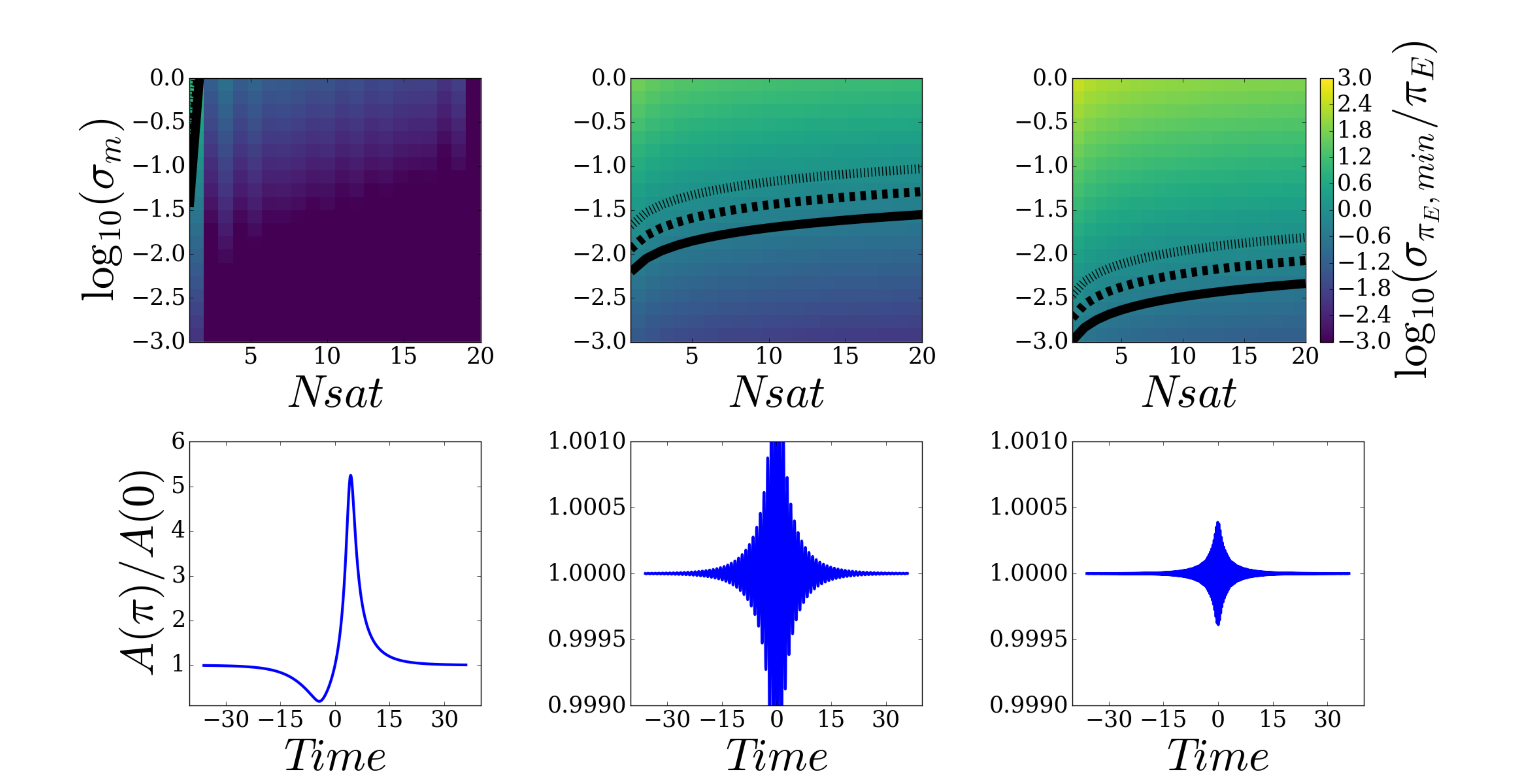}
    \caption{\textit{Top} : Minimum expected error on the parallax measurement $\sigma_{\pi_E,min}/\pi_E$ (color coded in $\log_{10}$ scale, in the range [-3,3]), assuming observing and lensing event details presented in the text. The small-dashed, dashed and solid contour lines indicate 1, 2 and 3 $\sigma$ detection regions. \textit{Bottom} : Magnification ratio between two satellites separated in phase by $\pi$. \textit{Left:} Solar orbit satellites. \textit{Midle:} Geosynchronous satellites.\textit{Right:} LEO satellites.}
    \label{fig:fleet}
\end{figure*}

We consider 
the fleet composed of $N_{sat} \in$ [1,20] spacecraft. To study the effect of varying telescope aperture 
we considered different photometric precision with $\sigma_m \in [0.1,1]$ mag. We assume 
$u_0=0.1$, $t_E=10 $ days, $\theta=45^{\circ}$ and $\lambda = 30 ^{\circ}$.
We distribute the fleet of telescopes equally in mean anomaly within the orbit. For example, in the case of three satellites, the phases are $0$, $2\pi/3$ and 
$4\pi/3$. We select an  observing window of 72 days around $t_0$ with a 1 hour cadence.
Since the Fisher information is additive, we simply sum the Fisher matrix of 
each satellite before the inversion to obtain the covariance matrix.
Results can be seen in the Figure~\ref{fig:fleet} and in the following we discuss details for various observing scenarios.

%%%%%%%%%%%%%%%%%%%%%%%%%%%%%%%%%%%%%%%%%%%%%%%%%%%%%%%%%%%%%%%%%%%%%%%%%%%%
\subsection{Fleet in solar orbit } \label{sec:SO}
%%%%%%%%%%%%%%%%%%%%%%%%%%%%%%%%%%%%%%%%%%%%%%%%%%%%%%%%%%%%%%%%%%%%%%%%%%%%
In this fleet configuration each space-telescope is orbiting the Sun at 1 AU. Then, the distance $d$ between two satellites is $d = 2\sin(\pi/N_{\rm{sat}})$ AU where $N_{\rm{sat}}$ is the total number of satellites (and assuming the telescope are evenly distributed on the orbit).  
  
The solar orbit present some advantages like low-cost thermal control. The main drawback 
is the distance with the Earth which seriously impact the required communications.
The advantage of such a configuration is the large orbital radius 
which produces large shifts $(\delta\tau,\delta\beta)$ in the various lightcurves. 
In fact, it is well known that the microlensing parallax is highly constrained 
with two observatories in this situation; it is the space-based parallax \citep{Refsdal1966, CalchiNovati2015,Street2016,Henderson2016}.
It is worth noting however that $\geq$ 5 telescopes with low precision (i.e. $\sigma_m\sim 1$ mag) can still strongly 
constrain the parallax, meaning that relatively small telescopes on inexpensive cube satellites could be a viable option.
%%%%%%%%%%%%%%%%%%%%%%%%%%%%%%%%%%%%%%%%%%%%%%%%%%%%%%%%%%%%%%%%%%%%%%%%%%%%
\subsection{Fleet in geosynchronous orbit } \label{sec:GO}
%%%%%%%%%%%%%%%%%%%%%%%%%%%%%%%%%%%%%%%%%%%%%%%%%%%%%%%%%%%%%%%%%%%%%%%%%%%%
A special orbit for a space-telescope is the geo-synchronous orbit in which the telescope stays above the same geographic location at a relatively large distance from Earth. This orbit has practical disadvantages as it is costly to reach and the risk of collision is comparatively high due to the existence of numerous commercial geosynchronous satellites.
We choose $R=42048$ km and a daily period for the simulations. If we assume that each observatory provides the same information $F_i$
to the parallax constrain, we can write:
\begin{equation}
F_{tot} \approx N_{sat} F_i
\end{equation}
where $F_{tot}$ is the total Fisher information. This directly leads to:
\begin{equation}
\sigma_{\pi_E}/\pi_E \propto \sigma_m/N_{sat}^{0.5}
\label{eq:error}
\end{equation}
We can rewrite  this equation and show that the required photometric precision to obtain a relative error on the parallax estimation $\delta=\sigma_{\pi_E}/\pi_E$ is:
\begin{equation}
\sigma_m \propto N_{sat}^{0.5}\delta
\label{eq:error}
\end{equation}
This trend is seen in the middle and right panels in Figure~\ref{fig:fleet}.  We can see that the parallax is well constrained if $\sigma_m\leq 0.01$ mag.
%%%%%%%%%%%%%%%%%%%%%%%%%%%%%%%%%%%%%%%%%%%%%%%%%%%%%%%%%%%%%%%%%%%%%%%%%%%%
\subsection{Low Earth Orbit } \label{sec:LOE}
%%%%%%%%%%%%%%%%%%%%%%%%%%%%%%%%%%%%%%%%%%%%%%%%%%%%%%%%%%%%%%%%%%%%%%%%%%%%
Space Agencies are more and more interested in the potential use of LEO satellite constellations. These constellations are extremely useful for simultaneous Earth observations. The benefits of this approach are multiple. One is the relative low-cost of orbital access. For example, the India Space Agency recently successfully released 104 small size satellites in a single mission\footnote{https://www.isro.gov.in/pslv-c37-successfully-launches-104-satellites-single-flight}, mostly tasked with Earth observations. It is also 
simple to use a Target of Opportunity (ToO) rapid-response mode, since communication with the satellites is relatively easy.
As shown by M16, a satellite in LEO is able to constrain the microlensing parallax, despite the relative low amplitude of the microlensing lightcurve's distortion due to the small orbital radius. \citet{Schvartzvald2016} obtained ToO observation from \textit{Swift} 
in order to constrain the parallax of the binary event OGLE-2015-BLG-1319. They showed that \textit{Swift} should have been able to constrain 
the parallax in principle. However, due to low sampling and low photometric precision, this was not the case for this event.
For this case, we selected $R=7000$ km (i.e $P\sim0.07$ days). From Figure~\ref{fig:fleet} right panel, it is clear that 
the parallax detection requires high photometric accuracy.

%%%%%%%%%%%%%%%%%%%%%%%%%%%%%%%%%%%%%%%%%%%%%%%%%%%%%%%%%%%%%%%%%%%%%%%%%%%%
\section{The real parallax detection efficiency} \label{sec:real}
%%%%%%%%%%%%%%%%%%%%%%%%%%%%%%%%%%%%%%%%%%%%%%%%%%%%%%%%%%%%%%%%%%%%%%%%%%%%
In the Section~\ref{sec:moon}, we have seen that a telescope placed on the Moon should be able to efficiently measure the parallax for the vast majority of microlensing events towards the Galactic Bulge. However, Section~\ref{sec:moon}, as well as G13 and M16, assumes that the model for an ongoing microlensing event is known. In fact, it is important to keep in mind that when an event is in progress, the microlensing model is usually not known. In other words, the Paczynski parameters (ie $t_0$, $u_0$ and $t_E$) need to be modeled at the same time as the parallax vector. This obviously adds complexity and one should expect that the theoretical results obtained in the previous section will be degraded. Moreover, the finite sampling and measurement precision directly lead to a fitted model different from the "true" model \citep{Bachelet2017}. M16 shows that :
\begin{equation}
\sigma_{\pi_E}/\pi_E \propto u_0^{0.5}
\label{eq:error}
\end{equation}
This clearly indicates that the parallax measurement depends on the $u_0$ fitted value.
Moreover, it is non trivial to select between different models based on real data. In practice, a $\Delta\chi^2$ is often used, using various thresholds to ensure a safe detection \citep{Yee2013}. In the present work, it is possible to use a more robust statistic, since we can simulate pure Gaussian errors. In this case, the Bayesian Information Criterion (BIC) is a efficient tool to distinguish real detections from overfitting, see for example \citet{Bachelet2012a,Bramich2016}. To illustrate this, we use the pyLIMA software package \citep{Bachelet2017} \footnote{https://github.com/ebachelet/pyLIMA} to simulate and model lightcurves corresponding to the Section~\ref{sec:moon}. We realize one fit with and one fit without the Moon parallax, and compute the $\Delta BIC$ for each events. 
\begin{figure}    
  \centering
  \includegraphics[width=9cm]{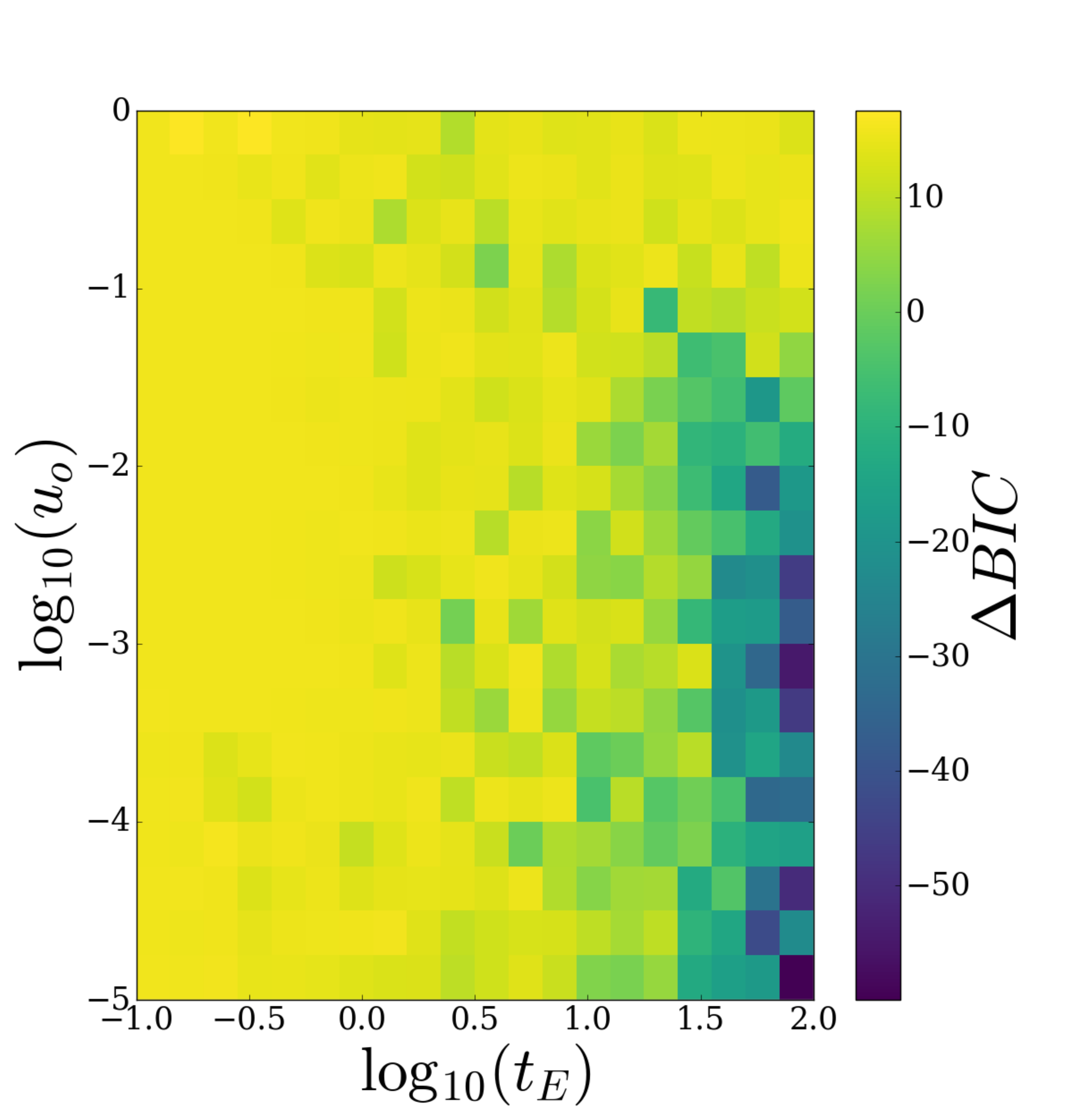}
    \caption{Parallax detection for microlensing events observed from the Moon. The positive detection region is reduce in comparison of the Figure~\ref{fig:Fisher1}.}
    \label{fig:reality}
\end{figure} 
As can be see in Figure~\ref{fig:reality}, the parallax detection is much harder than expected. All values with $\Delta BIC\sim20$  corresponds to $2\log(1454)$ (1454 is the number of data points for each lightcurve), and so corresponds to $\Delta\chi^2\sim0$. The reason is that the fitting process can slightly adjust the Paczynski parameters in order to fit the parallax. This problem is well known for the parallax constrain with ground data, see  Appendix~\ref{sec:annualparallax}.

%%%%%%%%%%%%%%%%%%%%%%%%%%%%%%%%%%%%%%%%%%%%%%%%%%%%%%%%%%%%%%%%%%%%%%%%%%%%
\section{Conclusions} \label{sec:conclusions}
%%%%%%%%%%%%%%%%%%%%%%%%%%%%%%%%%%%%%%%%%%%%%%%%%%%%%%%%%%%%%%%%%%%%%%%%%%%%

%%%%%%%%%%%%%%%%%%%%%%%%%%%%%%%%%%%%%%%%%%%%%%%%%%%%%%%%%%%%%%%%%%%%%%%%%%%%
We have studied the potential of various space observatories to systematically measure the microlensing parallax and hence to characterize the microlensing events. We first derive the exact Fisher matrix and compare our results to previous works. We then simulate various configurations corresponding to plausible future space missions. We show that the Moon is an ideal observatory to measure the parallax, assuming a  moderate photometric precision (0.01 mag). However, we moderate this conclusion in Section~\ref{sec:real}, since real observations require modeling and model selection, directly leading to a higher detection threshold. This is already well known for parallax measurement made with Earth observations (i.e the annual parallax) as discussed in the Appendix~\ref{sec:annualparallax}. We also simulate the potential of the WFIRST mission to detected the parallax on its own. We found that it is possible only for bright and long events (i.e $t_E>10$ days for a baseline photometric precision of $\sigma_m\sim0.001$ mag). Constellations of telescopes are promising. We confirm that telescopes orbiting the Sun at 1 AU have the strongest potential, as demonstrated in practice. However, both geosynchronous and low Earth orbits constellation are able to well constrain the parallax vector, assuming a sufficient number of satellite and/or good photometric precision since $\sigma_{\pi_E}/\pi_E \propto \sigma_mN_{sat}^{-0.5}$.
\section*{Acknowledgements}
The authors thank the anonymous referee for the constructive comments. This research has made use of NASA's Astrophysics Data System. Work by EB and RAS is support by the NASA grant NNX15AC97G. TCH acknowledges financial support from KASI grant 2017-1-830-03.

%%%%%%%%%%%%%%%%%%%%%%%%%%%%%%%%%%%%%%%%%%%%%%%%%%%%%%%%%%%%%%%%%%%%%%%%%%%%
%%%%%%%%%%%%%%%%%%%%%%%%%%%%%%%%%%%%%%%%%%%%%%%%%%%%%%%%%%%%%%%%%%%%%%%%%%%%
\appendix

%%%%%%%%%%%%%%%%%%%%%%%%%%%%%%%%%%%%%%%%%%%%%%%%%%%%%%%%%%%%%%%%%%%%%%%%%%%%
%%%%%%%%%%%%%%%%%%%%%%%%%%%%%%%%%%%%%%%%%%%%%%%%%%%%%%%%%%%%%%%%%%%%%%%%%%%%

\section{Fisher matrix analysis for the annual parallax} \label{sec:annualparallax}

\begin{figure}   [ht] 
  \centering
  \includegraphics[width=9cm]{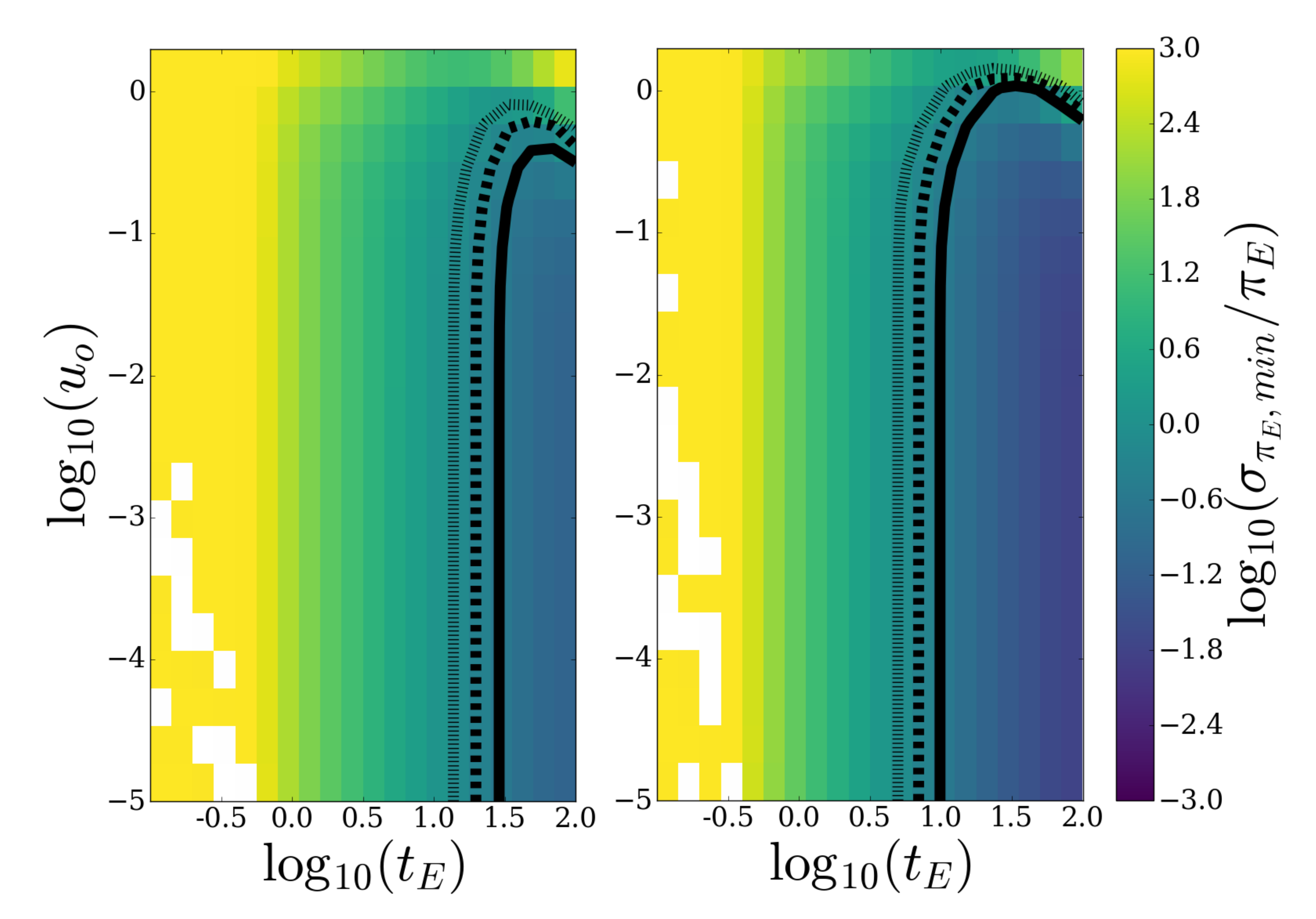}
    \caption{Similar to Figure~\ref{fig:Fisher1} for the annual parallax. \textit{Left:} Using 0.05 mag photometric precision. \textit{Right:} Using 0.01 mag photometric precision. Again, the blank pixels correspond to $\sigma^2_{\pi_E,min}(0)<0$ which is a signature of ill-observed event. }
    \label{fig:annual}
\end{figure}

%%%%%%%%%%%%%%%%%%%%%%%%%%%%%%%%%%%%%%%%%%%%%%%%%%%%%%%%%%%%%%%%%%%%%%%%%%%%
The annual parallax is the standard method used to measure the microlensing parallax. It is well known 
that such a measurement is in general possible only for long timescale events ($t_E>30$ days is a minimum). This is due
to the relatively long period and semi-major axis of the Earth's orbit around the Sun. Here we show that the Fisher matrix analysis can lead to overconfident conclusions. We conduct a similar study to that in Section~\ref{sec:moon} for the annual parallax, using the same simulation parameters, with the exception that $P=365.25$, $R=1$ AU, an observing window of 90 days around the event maximum and a one day cadence. We also simulate two baseline photometric precisions, namely 0.01 mag and 0.05 mag.
As can be seen in Figure~\ref{fig:annual}, the Fisher matrix analysis predicts that events with $t_E>15$ days should allow the systematic measurement of microlensing parallax, at least for the minimum photometric precision. However, it has been established from previous surveys that annual parallax measurements are extremely difficult for events with $t_E<30$ days, see for example \citet{Penny2016}.

%%%%%%%%%%%%%%%%%%%%%%%%%%%%%%%%%%%%%%%%%%%%%%%%%%%%%%%%%%%%%%%%%%%%%%%%%%%%

\section{Details of derivatives} \label{sec:derivatives}

%%%%%%%%%%%%%%%%%%%%%%%%%%%%%%%%%%%%%%%%%%%%%%%%%%%%%%%%%%%%%%%%%%%%%%%%%%%%
Here is the details of the model derivatives required for the Fisher matrix 
derivation.
%\begin{equation}
\begin{align*}
\frac{\partial A}{\partial u} &= {{-8}\over{u^2(u^2+4)^{3/2}}}  & \\ 
 \frac{\partial u}{\partial u_1} &= u_1/u  &  \frac{\partial u}{\partial u_2} &= u_2/u\\ 
\frac{\partial o_1}{\partial t_0} &= \omega\epsilon_\parallel\sin{\Omega} & \frac{\partial o_2}{\partial t_0} &= -\omega\epsilon_\bot\cos{\Omega} \\
\frac{\partial\delta\tau}{\partial t_0} &= \pi_E\cos{\theta}\frac{\partial o_1}{\partial t_0} + \pi_E\sin{\theta}\frac{\partial o_2}{\partial t_0} & \frac{\partial\delta\beta}{\partial t_0} &= -\pi_E\cos{\theta}\frac{\partial o_1}{\partial t_0} + \pi_E\sin{\theta}\frac{\partial o_2}{\partial t_0}\\
\frac{\partial u_1}{\partial t_0} &= (-1/t_E+\frac{\partial\delta\tau}{\partial t_0})\cos{\theta}-\frac{\partial\delta\beta}{\partial t_0}\sin{\theta} & \frac{\partial u_2}{\partial t_0} &= (-1/t_E+\frac{\partial\delta\tau}{\partial t_0})\sin{\theta}+\frac{\partial\delta\beta}{\partial t_0}\cos{\theta}\\
\frac{\partial u_1}{\partial u_0} &= \sin{\theta}  & \frac{\partial u_2}{\partial u_0} &= \cos{\theta}  \\
\frac{\partial u_1}{\partial t_E} &= -(t-t_0)/t_E^2\cos{\theta}  & \frac{\partial u_2}{\partial t_E} &= -(t-t_0)/t_E^2\sin{\theta} \\
\frac{\partial u_1}{\partial \pi_\parallel} &= o_1\cos{\theta}+o_2\sin{\theta} & \frac{\partial u_2}{\partial \pi_\parallel} &= o_1\sin{\theta}+o_2\cos{\theta}  \\
\frac{\partial u_1}{\partial \pi_\bot} &= -o_1\sin{\theta}+o_2\cos{\theta}  & \frac{\partial u_2}{\partial \pi_\bot}&= -o_1\cos{\theta}+o_2\sin{\theta}  \\
\end{align*}
%\end{equation}
\bibliography{biblio_moonpara.bbl}
\bibliographystyle{apj}
\end{document}